\documentclass[prb,preprint]{revtex4}
\usepackage{times}
\usepackage{amsmath}
\usepackage{amsfonts}
\usepackage{amssymb}
\usepackage{latexsym}
\usepackage{graphicx}
\usepackage{bm}
\usepackage{verbatim}

\begin{document}

\title{FCS and RICS Spectra of Probes in Complex Fluids}

\author{George D. J. Phillies}
\email{phillies@wpi.edu}

\affiliation{Department of Physics, Worcester Polytechnic
Institute,Worcester, MA 01609}

\date{\today}

\begin{abstract}
The fluorescence Correlation Spectroscopy (FCS) spectrum $G(t)$ and Raster Image Correlation Spectroscopy (RICS) spectrum $R(t)$ of dilute diffusing particles are determined by the displacement distribution function $P(x,t)$ of the particles and by the experimental parameters of the associated optical trains. This letter obtains the general relationships between $P(x,t)$ and these spectra.  For dilute diffusing molecules in simple liquids, $P(x,t)$ is a Gaussian in the displacement $x$; the corresponding $G(t)$ is a Lorentzian in $\sqrt{\langle (x(t))^{2}\rangle}$. In complex fluids such as polymer solutions, colloid and protein solutions, and the interior of living cells, $P(x,t)$ may have a non-Gaussian dependence on $x$, for example an exponential in $|x|$. We compare theoretical forms for FCS and for RICS spectra of two systems in which $P(x,t)$ is a Gaussian or an exponential in $x$, but in which the mean-square displacements are precisely equal at all times.  If the $G(t)$ and $R(t)$ arising from an exponential $P(x,t)$ are interpreted by using the forms for $G(t)$ and $R(t)$ that are appropriate for a Gaussian $P(x,t)$, the inferred diffusion coefficient may be substantially in error.
\end{abstract}

\maketitle

\section{Introduction}

This paper is a continuation of our previous work on fluorescence correlation spectroscopy studies of probes in complex fluids\cite{phillies2016a}. The previous paper considered systems in which relaxations were moderately non-exponential, so that the time correlation function $g^{(1)}(q,t) = \langle a_{q}(0) a_{q}(t)\rangle$ of a single spatial fourier component $a_{q}(t)$ of the fluorophore density at time $t$ could effectively be described by the first few terms of its time cumulant expansion.  While the cumulant expansion for $g^{(1)}(q,t)$ is always convergent, for severely non-exponential relaxations the cumulant expansion can become cumbersome.  This paper considers an alternative approach to treating fluorescence correlation spectroscopy (FCS) spectra, beginning with the distribution function $P(x,t)$ for a particle to diffuse a distance $x$ during time $t$.

Four decades ago\cite{magde1974a}, Fluorescence Correlation Spectroscopy (FCS) was identified as a general technique for measuring the diffusion coefficient and other properties of fluorescent and fluorescently-tagged molecules in solution.  In this technique, a volume of solution is illuminated with a focused laser beam.  The laser excites the fluorescent groups in the sample, causing them to emit fluorescent light. As the molecules diffuse through the illuminated region, the intensity of the fluorescence fluctuates in proportion to the number and positions of the fluorescent groups in the laser beam.  Experimentally, the time correlation function
\begin{equation}\label{eq:TCFFCS}
  G(t)  = \langle \mathcal{I}(\tau) \mathcal{I}(\tau+t)
\end{equation}
of the fluorescence intensity $\mathcal{I}(\tau)$ is determined, and used to calculate the diffusive properties of the fluorescing molecules.

$G(t)$ is determined by the intensity profile of the illuminating laser beam, by the collecting optical train, and by the displacement distribution function $P(x,t)$ of the diffusing molecules. In the original theoretical treatment of fluorescence correlation spectroscopy\cite{magde1974a}, which referred only to dilute solutions of fluorophores in simple Newtonian fluids,  $P(x,t)$ was taken to be a Gaussian
\begin{equation}
     \label{eq:gaussiandiffusion}
     P(x,t) = \frac{1}{(2 \pi \langle (x(t))^{2} \rangle)^{1/2}} \exp(- \frac{x^2}{2 \langle (x(t))^{2} \rangle})
\end{equation}
in which $\langle (x(t))^{2} \rangle$ is the mean-square distance a particle travels parallel to the $x$-axis during $t$. Equation \ref{eq:gaussiandiffusion} corresponds to the Langevin equation, as discussed in Berne and Pecora\cite{berne1976a}, Chapter 5;nthe form is aprpriate for the systems to which it was then applied. For simple Langevin-equation motion, $\langle (x(t))^{2} \rangle$ is related to the diffusion constant $D$ by
\begin{equation}\label{eq:meansquare}
   \langle (x(t))^{2} \rangle = 2 D t.
\end{equation}
As explained by Berne and Pecora\cite{berne1976a}, when the Langevin equation describes diffusion, then it is necessarily the case that $P(x,t)$ is a Gaussian in $x$, $\langle (x(t))^{2} \rangle$ increases linearly with time, $D$ is independent of time, and the quasielastic light scattering spectrum of the diffusing molecules is a single exponential in time.

At about the same time that FCS was developed, it was demonstrated for quasi-elastic light scattering spectroscopy (QELSS) that when the diffusing molecules are not dilute, the diffusion constant becomes the concentration-dependent diffusion coefficient\cite{phillies1974a}.  Furthermore, there are two physically-distinct translational diffusion coefficients\cite{phillies1974b}, the self diffusion coefficient and the mutual diffusion coefficient. For FCS, it was soon recognized\cite{phillies1975a} and then demonstrated experimentally\cite{scalettar1989a}  that if the diffusing molecules are not dilute, the diffusion coefficient being measured is determined by the fraction of the diffusing molecules that are fluorescently tagged.  If only a few of the diffusing molecules are tagged, FCS determines the molecular self-diffusion coefficient.  If all solute molecules are tagged, FCS determines the mutual diffusion coefficient. By way of comparison, quasi-elastic light scattering spectroscopy on a solution having one diffusing component always determines the mutual diffusion coefficient.

More recently, there has been interest in studying diffusion of probe molecules through complex fluids such as polymer solutions\cite{hallet1974a,hallett1976a}, protein solutions\cite{phillies1985a,phillies1985b}, and the interior of living cells\cite{luby1987a}.  In many of these systems, the matrix fluid surrounding the diffusing particles is viscoelastic rather than viscous.  Furthermore, the matrix fluid contains structures of various sizes.  As a result, the complex fluid cannot necessarily be approximated as being a viscoelastic continuum.  The Langevin model for diffusion is then inapplicable: The drag force on a diffusing macromolecule is not simply determined by its current velocity, and the so-called random thermal force on the diffusing particle may remain correlated with itself over extended periods of time.  For probes in complex fluids, Berne and Pecora's excellent Chapter 5 has nothing to say about probe diffusion. The discussion in Berne and Pecora's Chapter 11 provides a very general framework that might in principle be used to understand probe diffusion in complex fluids.

Piskorz and Ochab-Marcinek\cite{piskorz2014a} report an extensive Monte Carlo study to compute FCS spectra of particles performing restricted diffusion. They considered (i) a particle trapped by a harmonic potential whose center point itself diffuses, (ii) a particle free to move within an impenetrable spherical barrier whose center itself diffuses, and iii) a particle diffusing through a system that contains permeable barriers. The particle mean-square displacements in the three systems as functions of time were approximately equal.  They found that FCS spectra are substantially determined by particle mean-square displacements, the higher moments of $P(x,t)$ having little effect on $G(t)$. If one analyzes $G(t)$ for particles in spherical containments, or particles confronted with permeable barriers, by invoking the functional form for $G(t)$ for particles subjected to a diffusing harmonic potential, one obtains reasonably accurate values for the diffusion coefficient and the size of the confining volume. These three models of restricted diffusion are different, but they cannot readily be distinguished using FCS.

It has recently been recognized\cite{wang2009a} that in complex fluids $P(x,t)$ is not always a Gaussian. Wang, et al.,\cite{wang2009a} report the displacement distribution functions for colloidal beads diffusing along phospholipid bilayer tubes and for colloidal beads in concentrated actin solutions. Over a wide range of times, $P(x,t)$ for these systems is an exponential in $x$,
\begin{equation}
    \label{eq:granickdiffusion}
     P(x,t) = \frac{1}{L} \exp(-\frac{x}{L}),
\end{equation}
not the conventional Gaussian in $x$. Here $L$ is a range parameter. Wang, et al., found $L = a t^{1/2}$, $a$ again being a constant, which gives
\begin{equation}
    \label{eq:granickmeansquare}
    \langle (x(t))^{2} \rangle = 2 a^{2} t.
\end{equation}
Equation \ref{eq:granickmeansquare} is the Langevin-equation result for molecular diffusion, obtained under conditions in which the Langevin equation itself is very certainly not applicable. The non-Gaussian behaviors found by Wang, et al.\cite{wang2009a} are qualitatively very different from the features found by Piskorz and Ochab-Marcinek\cite{piskorz2014a}.  In restricted diffusion as studied by Piskorz and Ochab-Marcinek, $P(x,t)$ is heavily truncated beyond some containment distance. With an exponential $P(x,t)$, at shorter distances $P(x,t)$ is reduced relative to a Gaussian, but, at large $x$, $P(x,t)$ is much larger than a Gaussian having the same mean-square width.

We have previously explored several aspects of non-Gaussian diffusion and their effects on scattering and other methods of studying particle motion in complex fluids. Ref.\ \onlinecite{phillies2005a} treats dilute-particle diffusion, obtaining the relationship between the QELSS spectrum $S(q,t)$ and central moments of $P(x,t)$.  Ref.\ \onlinecite{phillies2012a} shows the additional terms that arise in $S(q,t)$ when the diffusing particles are not dilute.  Refs.\ \onlinecite{phillies2011a} and \onlinecite{phillies2015a} extend the analysis to consider particle motion as measured by pulsed-gradient spin-echo NMR. Refs.\ \onlinecite{phillies2013a} and \onlinecite{phillies2015b} reveal that experimental studies of probes diffusing through complex fluids have conclusively proven that the Gaussian diffusion approximation arising from the Langevin equation is generally invalid for probes in complex fluids.

Finally, we\cite{phillies2016a} calculated $G(t)$ for FCS spectra of probes in complex fluids in the form of an expansion in terms of the central moments $K_{n}(t)$ of $P(x,t)$. The central moment expansion of $P(x,t)$ is complete and convergent. Correspondingly, our expansion for $G(t)$ is complete.  However, for the non-Gaussian displacement distribution functions that correspond to non-exponential relaxations of QELSS spectra of the same systems, the convergence of the central moment expansion may be slow.  As a result, expressions for $G(t)$ based on the first few central moments of $P(x,t)$ may be less than satisfactory. This short paper therefore explores an alternative approach to computing $G(t)$, namely we obtain a general analytic form relating $G(t)$ to $P(x,t)$ and to the intensity profile of the illuminating beam, and then apply the form to systems\cite{wang2009a} for which $P(x,t)$ has been determined by direct experimental observation.

\section{Fluorescence Correlation Spectroscopy}

Our starting point is the general form $G(t)$ for the FCS time correlation function
\begin{equation}
     G(t) =  \int d{\bm r} d{\bm r'}  I({\bm r})  I({\bm r'}) P({\bm r'} -{\bm r} , t).
     \label{eq:generalform}
\end{equation}
 Here $I({\bm r})$ and $I({\bm r'})$ are the intensities of the illuminating laser beam at the points ${\bm r}$ and ${\bm r'}$, respectively, while $P({\bm r'} -{\bm r} , t)$ is the likelihood that a particle will move from ${\bm r}$ to ${\bm r'}$ during a time interval $t$.  The illuminating beams are taken to be cylinders, so that only motions in the $(x,y)$-plane, perpendicular to the beam axes, contribute to the time dependence of $G(t)$. The calculation then is effectively a two-dimensional problem. In the cases analysed here, $P({\bm r'} -{\bm r} , t)$ has translational invariance, so that it only depends on the displacement ${\bm R}$ between the start and finish points. In addition, the $x$ and $y$ components of the diffusive motion are independent, letting us write
\begin{equation}
       P({\bm r'} -{\bm r} , t) = P(x,t) P(y,t).
       \label{eq:factorP}
\end{equation}

Standard fourier transformation techniques allow us to replace the convolution integral of eq.\ \ref{eq:generalform} with a fourier-space integral
\begin{equation}
       G(t) = \frac{1}{2 \pi} \int d{\bm q} (I({\bm q}))^{2} F({\bm q}, t).
       \label{eq:generalformq}
\end{equation}

For a Gaussian-profile illuminating beam having a width $w$,
\begin{equation}
    I({\bm q}) = I_{o}  \exp(- q^{2} w^{2} / 2).
   \label{eq:gaussianbeam}
\end{equation}
The intermediate structure factor $F(({\bm q}),t)$ is
\begin{equation}
      F({\bm q},t) = \int dx  dy \exp(\imath q_{x} x + \imath q_{y} y)  P(x,t) P(y,t),
      \label{eq:ISF}
\end{equation}
where $q_{x}$ and $q_{y}$ are the $x$ and $y$ components of ${\bm q}$.

The fluorescence correlation function may up to constants be written
\begin{displaymath}
  G(t) = \int dq_{x} dq_{y} \exp(- q_{x}^{2} w^{2}) \exp(- q_{y}^{2} w^{2})
\end{displaymath}
\begin{equation}
     \times \int dx dy \exp(\imath q x) P(x,t)\exp(\imath q y) P(y,t).
     \label{eq:GTgeneral}
\end{equation}
or after rearrangement
\begin{displaymath}
  G(t) = \int dq_{x}dx \exp(- q_{x}^{2} w^{2})  \cos(q_{x} x) P(x,t)
\end{displaymath}
\begin{equation}
     \times \int dq_{y}  dy \exp(- q_{y}^{2} w^{2})  \cos(q_{y} y) P(y,t).
     \label{eq:GTgeneral2}
\end{equation}
The replacements of the complex exponentials with the cosines are permitted because $P(x,t)$ and $P(y,t)$ are even functions of $x$ and $y$.  The two lines of eq.\ \ref{eq:GTgeneral2} are the same except for a change of label. $P(x,t)$ is independent of $q$, so the integrals reduce to
\begin{equation}
      G(t) = \left( \int dq_{x} dx \exp(- q_{x}^{2} w^{2})  \cos(q x) P(x,t) \right)^{2}
      \label{eq:Gtgeneralfinal}
\end{equation}
as the general form for the FCS time correlation function in terms of the displacement distribution function. On performing the integral over $q_{x}$, we obtain
\begin{equation}\label{eq:kernelform}
       G(t)  = \left(\int dx  W(x) P(x,t)\right)^{2}
\end{equation}
with
\begin{equation}\label{eq:Wxform}
  W(x) = (2 \pi w^{2})^{-1/2} \exp( -\frac{x^{2}}{4 w^2})
\end{equation}

In a single experiment $w$ is a constant, much as the scattering vector ${\bm q}$ is a constant in a single quasielastic light scattering spectroscopy experiment. The weighting function $W(x)$ causes the experiment to sample a Gaussian-weighted central sample of $P(x,t)$.  $P(x,t)$ has a domain having width $S$ within which it is significantly non-zero.  $S$ increases with increasing $t$.  At small $t$, $S \ll w$.  $S$ increases, but $P(x,t)$ is substantially non-zero only in narrow regions within which $W(x)$ is nearly constant, so $G(t$) is nearly constant.   At large $t$, $S \gg w$, so $G(t)$ is determined by the central core of $P(x,t)$.  At large $t$, the behavior of $P(x,t)$ in its wings does not contribute to $G(t)$, because $W(x)$ is nearly zero for large $x$. However, in many cases $P(x,t)$ deviates most prominently from simple Gaussian behavior in its wings, so $G(t)$ can readily be insensitive to non-Gaussian behavior. To use FCS to examine non-Gaussian behavior at a particular time $\tau$, one needs to choose $w$ such that $x/w \approx 1$ for $x$ in the region where non-Gaussian behavior occurs near time $\tau$.  This choice of $w$ may be inauspicious for observing the behavior of $G(t)$ at other times.

We now evaluate eq.\ \ref{eq:Gtgeneralfinal} for two cases of $P(x,t)$, namely the Gaussian case corresponding to Langevin-equation diffusion and the pure-exponential case found by Wang, et al.\cite{wang2009a}.  The Gaussian case was first evaluated by Magde, et al.\cite{magde1974a}.  We write the mean-square displacement in one dimension as $\langle (x(t))^{2} \rangle = 2 a^{2} t$, $a$ being a constant. For Gaussian diffusion
\begin{equation}
     \label{eq:FCSgaussian}
      G(t) = \frac{1}{2 \pi w^{2}(1+ \frac{a^2 t}{w^{2}})},
\end{equation}
which is a Lorentzian function in the variable $w$. For exponential diffusion, one finds
\begin{equation}\label{eq:FCSexponential}
      G(t) = \frac{1}{ a^{2}t} \exp(\frac{w^{2}}{a^{2}t}) \left( {\rm Erf}
      \left(\frac{w}{\sqrt{2a^{2}t}}     \right)-1 \right)^{2}.
\end{equation}

If one has a system in which eq.\ \ref{eq:FCSexponential} is correct, and attempts to interpret $G(t)$ by fitting it to an  expression for Gaussian diffusion, namely
\begin{equation}\label{eq:fittingform}
        G(t) = \frac{h}{4 \pi (w^{2} + a^{2} t)},
\end{equation}
the outcome may be misleading. Here we have taken not only $a$ but also the amplitude $h$ to be free parameters. In making the following numerical fits, $G(t)$ was always calculated from very nearly $t=0$ to times such that $G(t)$ had decayed through two orders of magnitude.

For example, suppose $a=1$, so the mean-square displacement is unity at $t=1$, and suppose the illuminating beam has unit width, $w=1$. If one uses eq.\ \ref{eq:FCSexponential} and these parameters to compute $G(t)$, and then fits the computed $G(t)$ to eq.\ \ref{eq:fittingform}, the outcome of the fit is not the correct $a=1$ but instead $a=0.62$. The diffusion coefficient $D$ inferred from $\langle (x(t))^{2} \rangle = 2 a^{2} t = 2 D t$ is then in error by nearly a factor of three.  As seen in Table One, if $P(x,t)$ were actually exponential, but $G(t)$ was interpreted assuming a Gaussian $P(x,t)$, for these parameters $a$ from the fitting process would consistently be 0.62 of its correct value, for a wide range of correct values of $a$.

We emphasize that in eq.\ \ref{eq:fittingform} the zero-time amplitude $h$ was taken to be a free parameter, as opposed to forcing $h = 1$.  With $w = 1$ and the actual $a=1$, if $h=1$ had been forced during the fitting procedure, a nonlinear least-squares fit leads to $a=0.267$, leading to a fifteen-fold error in the inferred diffusion coefficient.  It should not be assumed from this single numerical test that treating $h$ as a free parameter rather than forcing $h=1$ will always lead to a less wrong value for $a$.

\begin{center}
\vspace*{1em}
\begin{table}[t!]
  \begin{tabular}{|l|l|}
  \hline
$A$ & $a$ \\
0.1 &0.0639\\
0.3 & 0.192\\
1.0 & 0.623\\
3.0 & 1.898\\
\hline
\end{tabular}
  \caption{Inferred value of $a$ for a particle having an exponential displacement distribution function (eq.\ \ref{eq:granickdiffusion}) with $L = A t^{1/2}$ and unit $w=1$ beam width. Observe $a \approx 0.63 A$ throughout.}\label{table:tableone}
\end{table}
\end{center}

\section{Raster Image Correlation Spectroscopy}

Raster Image Correlation Spectroscopy (RICS) is a variant on FCS in which several different locations in the same sample are illuminated.  The fluorescent intensities at different locations and times are then cross-correlated to study diffusion\cite{digman2005a,digman2005b}.  The term 'raster' is used because the illuminated positions in the original experiments lay on a rectangular grid.  The two illuminating beams are again taken to have Gaussian beam profiles, but their centers are displaced from each other by ${\bm b}$.  In the following, the vector ${\bm b}$ is taken to lie along the $x$-axis, which may or may not be parallel to one of the raster axes.  At this point the calculation differs from the results in refs.\ \onlinecite{digman2005a} and \onlinecite{digman2005b}, in which the $x$ and $y$ axes were taken to be the raster axes. Also, here ${\bm b}$ is treated as a continuous variable. The cross-correlation function is
\begin{equation}\label{eq:RICS1}
   R(t) =   \langle \mathcal{I}({\bm r}, \tau) \mathcal{I}({\bm b} + {\bm r'}, \tau + t)\rangle,
\end{equation}
which may be written
\begin{equation}\label{eq:ricspositionaverage}
      R(t) = \int d{\bm r} \int d{\bm r'} I({\bm r}) I({\bm r'}) P({\bm r'}+{\bm b}-{\bm r}, t).
\end{equation}
The origins of ${\bm r}$ and ${\bm r'}$ are the centers of the two illuminating laser beams. The steps that led to eq.\ \ref{eq:Gtgeneralfinal} now lead instead to
\begin{displaymath}
     R(t) =  \left( \int_{-\infty}^{\infty} dy \int_{-\infty}^{\infty} dq_{y}
     \exp(- q_{y}^{2} w^{2}/2) \exp(- q_{y}^{2} w^{2}/2)  \cos(q_{y} y) P(y,t) \right)
\end{displaymath}
\begin{equation}\label{eq:ricsdxdy}
      \times  \left( \int_{-\infty}^{\infty} dx  \int_{-\infty}^{\infty} dq_{x}
      \exp(- q_{x}^{2} w^{2}/2)   \exp(- q_{x}^{2} w^{2}/2) \cos(q_{x} x) \cos(q_{x} b) P(x,t) \right).
\end{equation}
The first line refers to motion perpendicular to the ${\bm b}$ axis, so it is the same as the kernel of eq.\ \ref{eq:Gtgeneralfinal}, while the second line refers to motion parallel to the ${\bm b}$ axis. After integrating on $q_{x}$ and $q_{y}$, eq.\ \ref{eq:ricsdxdy} factors into
\begin{equation}
    R_{x}(t) = \int_{-\infty}^{\infty} dx (16 \pi w^{2})^{-1/2} \exp\left(-\frac{(x+b)^{2}}{4 w^{2}}\right)\left(1 + \exp\left(\frac{b x}{w^{2}}\right) \right) P(x,t),
    \label{eq:Rxt1}
\end{equation}
and
\begin{equation}\label{Ryt1}
   R_{y}(t) = \int_{-\infty}^{\infty} dy  \left(\frac{1}{4 \pi w^{2}}\right) \exp\left(- \frac{y^{2}}{4 w^{2}}.    \right)  P(y,t)
\end{equation}
     $R_{y}(t)$ was evaluated in the previous Section.

For the Gaussian $P(x,t)$ of eq.\ \ref{eq:gaussiandiffusion},
\begin{equation}
       \label{eq:ricsgaussian}
       R_{x}(t) = \frac{\exp \left(-\frac{b^{2}}{4 (a^{2} t + w^{2})} \right)}{\sqrt{4 \pi (a^{2}t + w^{2}) }}
\end{equation}
while for the exponential $P(x,t)$ of eq.\ \ref{eq:granickdiffusion},
\begin{displaymath}
      R_{x}(t) = \frac{1}{2 a \sqrt{t}} \exp\left(-\frac{b^{2}}{4 w^{2}}\right)  \left[
      -\exp\left(\frac{(a b \sqrt{t} -2w^{2})^{2}}{4 a^{2} t w^{2}} \right)
      \left( -1 + {\rm erf}\left[\left(\frac{w}{a \sqrt{t}} - \frac{b}{2w}\right)\right]   \right)\right.
\end{displaymath}
\begin{equation}
     \left. - \exp\left(\frac{( a b \sqrt{t} +2 w^{2})^{2}}{(4 a^{2} t w^{2})}\right)
       \left( -1 + {\rm erf}\left[\frac{w}{a \sqrt{t}} + \frac{b}{2w} \right]   \right)\right).
      \label{eq:ricsexponential}
\end{equation}
$R(t)$ is obtained by multiplying each $R_{x}(t)$ by its corresponding $R_{y}(t)$.

As an illustration of the effect of assuming that $P(x,t)$ has a Gaussian form, when $P(x,t)$ is in fact exponential, we take the relaxation function $R(t)$ for an exponential $P(x,t)$ and attempt to extract $a$ from it by fitting it to the $R(t)$ appropriate for a Gaussian $P(x,t)$. We choose $a=1$, take beam diameter $w=1$, and determine what $a$ is obtained from the fit, as a function of the displacement $b$.  If the fit were valid, we would obtain $a=1$ throughout.  We in fact find the results seen in Table 2.  There is a weak dependence of $a$ on $|{\bm b}|$, but the fitted $a$ is incorrect throughout.
\begin{center}
\vspace*{1em}
\begin{table}[t!]
  \begin{tabular}{|l|l|}
  \hline
$b$ & $a$ \\
12 & 0.779\\
7 & 0.712\\
3 &0.715\\
1 &0.670\\
0.1 & 0.674\\
\hline
\end{tabular}
  \caption{Inferred value of $a$ from the RICS spectrum for various choices of the displacement $b$, for a particle having an exponential displacement distribution function (eq.\ \ref{eq:granickdiffusion}) with $L = A t^{1/2}$ and the actual $A$ equalling 1.   }\label{table:tabletwo}
\end{table}
\end{center}

\section{Discussion}

For simple Gaussian diffusion, the FCS spectrum has the form
\begin{equation}
       G(t) = A (1+ \frac{D t}{w^{2}})^{-1},
       \label{eq:gtaulangevin2}
\end{equation}
implying a natural time $\tau_{D}=w^{2}/D$. Here $A$ is a constant.

If diffusion is not Gaussian, the dependence of $G(t)$ on its parameters changes. The FCS spectrum may depend on beam diameter $w$ in ways other than the one seen in eq.\ \ref{eq:gtaulangevin2}. Is such a dependence of $G(t)$ on beam diameter $w$ purely hypothetical, or can such dependences be observed? Experiments showing that the dependence of $G(t)$ on $w$ can deviate from eq.\ \ref{eq:gtaulangevin2} have already been performed.  Note results of  Wawrezinieck, et al.\cite{wawrezinieck2005a} and Masuda, et al.\cite{masuda2005a}. These studies varied $w$ by nearly a factor of 2. Wawrezinieck, et al.\cite{wawrezinieck2005a}, studied a labelled protein and a labelled lipid in COS-7 cells. For both labelled species, $\tau_{D}$ was linear in $w^{2}$.  However, if one extrapolated $\tau_{D}$ to its $w \rightarrow 0$ limit, one would find $\tau_{D} < 0$ or $\tau_{D} > 0$, respectively, for the two probes.  When the beam diameter is reduced to zero, the residence time of labelled molecules in the beam is obliged to fall to zero, implying that $\tau_{D}$ for these systems is not linear in $w^{2}$, if only in the range of small $w$ where $G(t)$ was not observed. Masuda, et al.\cite{masuda2005a}, studied the diffusion of a small molecule through hyaluronan solutions.  In dilute solution, changing the beam diameter had almost no effect on the inferred diffusion coefficient. In non-dilute hyaluron solutions, the inferred $D$ depended on $w$. Masuda, et al., interpreted their measurements as indicating that the polymer was more effective at hindering $D$ over larger distances than over shorter distances.

The anomalous effects found by Wawrezinieck, et al., and Masuda, et al., could not have arisen if $P(x,t)$ had been a Gaussian in their systems. If $P(x,t)$ is a Gaussian in $x$, it is necessarily the case that $\tau_{D} = w^{2}/\langle (x(t))^{2} \rangle$.  The mean-square displacement $\langle (x(t))^{2} \rangle$ is a property of the fluid and is entirely independent of the beam diameter $w$, so $\tau_{D} \sim w^{2}$.  For a Gaussian $P(x,t)$, changing $w$ changes the extent to which a given $\langle (x(t))^{2} \rangle$ leads to a change in $G(t)$, but does not affect the value of $\langle (x(t))^{2} \rangle$ to be inferred from the observed spectrum

For a non-Gaussian $P(x,t)$, $G(t)$ depends on the full shape of $P(x,t)$, not just on its second moment, so a fit that assumes that $P(x,t)$ is entirely determined by $\langle (x(t))^{2} \rangle$ sometimes leads to invalid results, as seen in Table 1 for the case of an exponential $G(t)$. On the other hand, simulations indicate\cite{piskorz2014a} that in some interesting cases the error from assuming Gaussian diffusion is small.

Several interesting analogies appear here with quasielastic light scattering spectroscopy.  First, the beam diameter $w$ of FCS is directly analogous to the scattering wavevector $q$ of QELSS.  $w$ and $q$ determine, albeit in different ways, the displacement distances to which the experiment is sensitive.  Just as the QELSS spectrum is properly described as $S(q,t)$ and not $S(t)$, so also the FCS spectrum is properly described as $G(w,t)$ and not $G(t)$.  Second, as shown by simulations of Piskorz and Ochab-Marcinek\cite{piskorz2014a} it is sometimes the case that deviations from Gaussian behavior do not have a large effect on the diffusion coefficient inferred from an FCS spectrum.  Correspondingly, while there are theoretical conditions under which $S(q,t)$ would have a significant $q$-dependence, obtaining such conditions experimentally proved historically to be a significant challenge. There is a direct analogy with measuring $D$ using quasi-elastic light scattering, in which measuring the QELSS spectrum $S(q,t)$ for several values of $q$ tests for deviations from simple diffusive behavior. Correspondingly, for FCS and RICS, varying the distances to which the experiment is sensitive (varying $w$ or $q$) can be a useful test of the validity of the measurement.

\end{document}